# Deconvoluting Reversal Modes in Exchange Biased Nanodots

Randy K. Dumas,[1] Chang-Peng Li,[2] Igor V. Roshchin,[3,4] Ivan K. Schuller[2] and Kai Liu[1,*]

[1]*Physics Department, University of California, Davis, California 95616*
[2]*Physics Department and Center for Advanced Nanoscience, University of California - San Diego, La Jolla, California 92093*
[3]*Department of Physics and Astronomy, Texas A&M University, College Station, Texas, 77843*
[4]*Materials Science and Engineering Program, Texas A&M University, College Station, Texas, 77843*

**Abstract**

Ensemble-averaged exchange bias in arrays of Fe/FeF$_2$ nanodots has been deconvoluted into local, microscopic, bias separately experienced by nanodots going through different reversal modes. The relative fraction of dots in each mode can be modified by exchange bias. Single domain dots exhibit a simple loop shift, while vortex state dots have asymmetric shifts in the vortex nucleation and annihilation fields, manifesting local incomplete domain walls in these nanodots as magnetic vortices with tilted cores.

Ferromagnet/antiferromagnet (FM/AF) exchange biased nanostructures[1] have critical applications in spin-valve type of spintronic devices.[2] They have also excited recent interests in antiferromagnetic spin-transfer torque effect,[3, 4] low-power electrically controlled magnetic switching in multiferroics,[5-7] and exchange biased magnetic vortices,[8-11] which have novel potential applications in nanoelectronics and magnetic memory. Such nanostructures not only have structural heterogeneity across the FM/AF interface, but also often have nanoscale confinements in the lateral dimension,[12-15] leading to intriguing ground state domain configurations and reversal mechanisms that are yet to be fully understood.

The exchange bias (EB) in these systems is conventionally obtained from the shift of the major hysteresis loop at the coercive fields, defined as $H_E$. However, some systems show very different unidirectional anisotropy energies when measured by reversible vs. irreversible methods,[16, 17] and the anisotropy may be dependent upon sample *lateral* dimensions. For instance, in NiO/NiFe and FeMn/NiFe microstructures the $H_E$ is larger than in uniform films;[18] a size-dependence of EB is recently reported in Co/CoO nanostructures.[19] These results suggest an $H_E$ distribution in seemingly uniform films, raising the question whether the $H_E$ obtained from the major loop is uniform across the entire sample or just a reflection of the weakest point. In prior studies of Fe/epitaxial-FeF$_2$ films a distribution of $H_E$ was shown to be centered on the major loop value.[20] However, since the FM layer was continuous, variations of exchange coupling within the layer may mask the distribution of local EB. *A clear link is still missing between the ensemble-averaged $H_E$ measured by magnetometry and the local, microscopic, exchange bias*. In addition, local EB in FM/AF hybrids can be tailored by nano-patterning.[12-14] In particular, the local balance between magnetostatic and exchange energies in the FM can strongly modify the magnetization reversal mode, e.g., single domain (SD) vs. vortex state (VS) reversal. The interplay between exchange bias and the vortex state reversal mode has led to a host of fascinating properties, including angular dependent magnetization reversal,[8] chirality control,[10] and vortex imprinting into the AF.[9, 11]

Furthermore, depth-dependent magnetic configurations across the FM/AF interface are also important. Recent studies have highlighted the role of local incomplete domain walls (LIDWs) in continuous films in generating the magnetization reversal asymmetry commonly observed.[21, 22] Such LIDWs form parallel to the FM/AF interface due to the competition between inhomogeneous interfacial exchange and the magnetic field, and their lateral extent can change within the film plane under field cycling. However, the existence and behavior of such LIDWs in EB nanostructures with confined lateral dimensions have remained unexplored.

In this work we report the interplay between geometric confinement and exchange bias in patterned arrays of Fe/FeF$_2$ nanodots of sizes close to the SD - VS reversal mode boundary for Fe. The major loop EB has been deconvoluted into weighted averages of local EB separately experienced by SD and VS dots. The EB also modifies the relative fraction of dots that undergo VS and SD reversal modes. These findings establish the long missing link between macroscopic and local EB. Furthermore, we have observed unequal shifts in the bias manifested by the VS nucleation and annihilation fields, which provides first experimental demonstration of LIDWs in patterned structures as magnetic vortices with tilted cores.

Polycrystalline Fe (20 nm) / FeF$_2$ (50 nm)/Ag cap (5 nm) nanodots were grown on Si substrates using nanoporous alumina shadow masks in conjunction with electron beam evaporation and Ar-ion etching.[23] The nanodot size is 67±13 nm, and coexistence of SD and VS modes is expected from our prior studies on unbiased Fe dots.[24, 25] The dot center-to-center spacing is roughly twice the dot diameter, as illustrated in Fig. 1 inset, so that the dipolar interactions between the dots are negligible.[26]

Magnetic properties were measured in a field applied in the plane of the nanodots using a vibrating sample magnetometer with a liquid helium flow cryostat. Major hysteresis loops were measured at room temperature (RT) and at 40 K, after two different cooling procedures. The field-cooling (FC) procedure involves cooling from RT through the FeF$_2$ Néel temperature (T$_N$=78 K) in a 5 kOe applied field to 40 K. Subsequent measurements were then taken with the applied field parallel to the FC direction. The zero-field-cooling (ZFC) procedure involves ac demagnetizing the sample at RT and then cooling in zero applied field to 40 K. Note that no training effect was observed and the field sweep rate (~12 Oe/s) was slow enough to avoid any relaxation processes.

The first-order reversal curve (FORC) technique[24, 25, 27] was employed to investigate details of the magnetization reversal. The FORC distribution is defined as

$$\rho(H, H_R) \equiv -\frac{1}{2}\frac{\partial^2 M(H, H_R)/M_S}{\partial H \partial H_R}, \qquad (1)$$

where $H_R$ is the reversal field, $H$ is the applied field, $M$ is the magnetization, and $M_S$ is the saturation magnetization. This distribution eliminates the purely reversible components of the magnetization. Thus any non-zero $\rho$ corresponds to *irreversible* switching processes. The FORC distribution can also be plotted in coordinates of ($H_C$, $H_B$), where $H_C=(H-H_R)/2$ is the local coercive field and $H_B=(H+H_R)/2$ is the local interaction or bias field. The integration

$$I_{irrev} = \int \rho(H_C, H_B) dH_C dH_B \qquad (2)$$

captures the total amount of irreversible magnetic switching.[20] A quantitative measure of the relative fraction of dots with a particular reversal mode is evaluated by selectively integrating the FORC distribution over the region of interest.

Magnetic hysteresis loops at 40 K are shown in Fig. 1. The FC sample displays a shifted loop and a coercivity enhancement from 348 Oe at RT to 456 Oe. Conventionally, $H_E$ is determined from the major loop (ML) shift according to

$$H_E^{ML} = (H_C^R + H_C^L)/2 \tag{3}$$

where $H_C^R$ and $H_C^L$ are the right and left coercive fields, respectively. Using this, an exchange bias of $H_E^{ML} = -97$ Oe is found. Note that compared to Fe/epitaxial-FeF$_2$ films,[20] these nanodots exhibit an enhanced coercivity caused by the nano-patterning,[16] and a reduced EB due to the polycrystalline nature of the FeF$_2$.[28] In contrast, the ZFC sample shows no EB, but an increased coercivity of 447 Oe at 40K. The coercivity increase from RT to 40K in the FC and ZFC samples is due to both the influence of the FeF$_2$ and the reduced temperature.[25]

The RT FORC's and the corresponding FORC distribution in the ($H$, $H_R$) coordinate system are shown in Figs. 2(a) and 2(b), respectively. The FORC distribution is similar to that observed previously in comparable sized and unbiased Fe nanodot arrays, displaying a SD and VS reversal mode mixture.[24, 25, 29] In Fig. 2(b) the FORC feature highlighted by the dashed box corresponds to the SD reversal mode; the other three peaks highlighted by circles correspond to the VS reversal mode. By projecting the former FORC distribution $\rho$ along the bias field $H_B$ axis, the EB of the SD dots can be determined. As shown in Fig. 3, for SD dots, the $dM/dH_B$ peak at RT is centered on $H_B=0$, consistent with an EB of $H_E^{SD} = 0$ Oe. Additionally, by selectively integrating the SD region of the FORC distribution, and using Equation (2), the fraction of dots reversing at RT by a SD reversal mode, $f_{SD} = 27 \pm 2\%$.[30]

For the VS dots, FORC peaks and their widths are directly related to the VS nucleation/annihilation fields and their corresponding distributions. A schematic VS hysteresis loop is shown in Fig. 2(e). Along the descending (ascending) field branch, a vortex nucleates at $H_{N1}$ ($H_{N2}$) and annihilates at $H_{A1}$ ($H_{A2}$); as the field is reversed to positive saturation immediately after $H_{N1}$, the annihilation field is $H_{A3}$ (dashed line in Fig. 2(e)). Fig. 2(f) is a schematic FORC distribution in the ($H$, $H_R$) coordinate system highlighting the relationship between the three primary FORC peaks and the nucleation/annihilation fields shown in Fig. 2(e). For the RT FORC distribution shown in Fig. 2(b), along the descending branch, the VS nucleation field ($H_{N1}$ = 140 Oe) and annihilation field ($H_{A1}$ = -1,340 Oe) are symmetric to those along the ascending branch ($H_{N2}$ = -140 Oe, $H_{A2}$ = 1,340 Oe).[31] Thus, the dots that undergo VS reversal do not exhibit EB at RT either, as expected.

After FC the sample, the 40 K FORC's and the corresponding distribution are shown in Figs. 2(c) and 2(d), respectively. Although they appear qualitatively similar to those at RT, the peak positions and relative intensities in the FORC distribution are quite different. By integrating over the SD feature we find the fraction of SD dots has increased to $f_{SD}$ = 32±3 %. Additionally, as shown in Fig. 3, the SD peak is shifted towards negative bias fields. The projection of $\rho$ ($H_C$, $H_B$) along the $H_B$-axis, $dM/dH_B$, peaks at -50 Oe, indicating an exchange bias of $H_E^{SD} = -50$ Oe. Interestingly, this value is nearly half the ML value, $H_E^{ML} = -97$ Oe shown in Fig. 1. Note that $H_E^{ML}$ effectively measures the shift in magnetization switching fields along the descending and ascending branches. Equivalently, and in analogy to Eqn. 3, an exchange bias can be calculated from the shift of the VS nucleation fields as follows:

$$H_E^{VS,N} = (H_{N1} + H_{N2})/2. \tag{4}$$

From the FORC distribution in Fig. 2(d), we find $H_{N1}$ = -265 Oe and $H_{N2}$ = -13 Oe, hence $H_E^{VS,N}$ = -139 Oe, significantly larger than the ML value. This key difference illustrates that SD and VS dots are affected by exchange bias differently for the same FC procedure. As the ML is an ensemble average of both SD and VS phases, we extract an "averaged" EB using the calculated exchange biases and phase fractions from the FORC analysis:

$$H_E^{FORC} = H_E^{SD} \times f_{SD} + H_E^{VS,N} \times (1 - f_{SD}) \quad (5)$$

This weight averaged exchange bias is -110 Oe, similar to the value found from the ML analysis. Thus, the conventionally determined $H_E$ can be deconvoluted into distributions of local exchange bias experienced by all the dots in the assembly.

Additionally, the FORC analysis of the VS reversal reveals a much more interesting and complex behavior than a simple shift of the major loop would suggest. Analogous to Eqns. 3 and 4, an exchange bias can be calculated from the shift of the VS annihilation fields:

$$H_E^{VS,A} = (H_{A1} + H_{A2})/2 \quad (6)$$

Using the VS peak positions in Fig. 2(d), we find $H_{A1}$ = -1,523 Oe and $H_{A2}$ = +1,400 Oe, thus $H_E^{VS,A}$ = - 61 Oe. This is significantly different from the major loop bias, and much smaller than the shift of the nucleation fields $H_E^{VS,N}$ calculated using Eqn. 4. If the VS loop was simply shifted along the applied field axis, the values calculated using Eqns. 4 and 6 should be identical, i.e., both $H_N$ and $H_A$ are displaced by the same amount. However, the FORC results indicate that the VS nucleation fields are experiencing a larger displacement than the annihilation fields. This results in an asymmetric VS major loop, as schematically shown in the inset of Fig. 2(e), where the top portion of the loop is wider than the bottom.

Such an asymmetry is a signature of the LIDWs[21, 22] which have spring-magnet[32] like varying magnetization along the depth of the FM, and are manifested in nanodots as magnetic vortices with tilted cores. The Monte Carlo simulations of Mejía-López *et al*[33] model a nearly identical material system as the Fe/FeF$_2$ dots studied here. They find a pronounced asymmetric reversal of the VS major loop due to a non-uniform magnetization through the thickness of the Fe layer. In fact, their simulated hysteresis loops are qualitatively similar to Fig. 2(e) inset, which is representative of the $H_N$ and $H_A$ values found after FC the sample. The analytical calculations of Guslienko and Hoffmann[34] also confirm the type of asymmetry we observe. In their calculations the asymmetry is dictated by the quantity $\ln(R/R_c)/2$, where $R$ is the radius of the dot and $R_c$ the radius of the vortex core. Using values appropriate for this sample, $R$=33.5 nm and $R_c$=8 nm:[35]

$$\ln (R/R_C )/2 < 1 \rightarrow |H_E^A| < |H_E^N| \qquad (7)$$

That is, the apparent shift of the nucleation fields is larger than that of the annihilation fields.[36] This happens because in relatively small dots, where the vortex core has an energy contribution comparable to that of the rest of the dot, the energy barrier for the vortex nucleation (which requires an area with out-of-plane magnetization) is much larger than the barrier for annihilation. Such a difference between the shifts of the nucleation and annihilation fields is indeed what we have observed experimentally. Just as in the Monte Carlo simulations, the observed asymmetry is a result of depth dependent magnetization variations in the FM. In other words, the degree of pinning induced by the AF is stronger near the FM/AF interface than at the free surface of the FM, leading to the formation of LIDWs and magnetization reversal asymmetry.[22] The unidirectional magnetic anisotropy induced by the EB suppresses the VS. This suppression results in a delayed vortex nucleation and earlier vortex annihilation at the FM/AF interface during reversal, as compared to the free FM surface. As the lateral extent of such LIDWs is limited by the dot size, and a vortex core exists in the dot, the non-collinear magnetic moments along the depth of the FM manifest as a tilted vortex core.

Finally, we examine the evolution of the SD phase fraction. As mentioned above, the SD fraction $f_{SD}$ increases from 27±2% at RT to 32±3% in the FC sample at 40 K. As previously studied in unbiased Fe dots,[25] $f_{SD}$ can simply increase at lower temperatures due to suppression of thermally assisted vortex nucleation/annihilation. In order to distinguish the influence of EB alone on the SD phase fraction, we have investigated nanodots at 40 K under ZFC. From the FORC analysis we find that the projection of the SD reversal along the $H_B$-axis (Fig. 3) and VS peaks show no EB. However, integration of the FORC diagram shows that after ZFC, $f_{SD}$ is only 28±3%, which is smaller than that after FC and nearly the original RT value. After demagnetizing the sample at RT, most of the dots are in the VS. Upon ZFC to 40 K, while the lower temperature alone should increase $f_{SD}$, there is a competing mechanism which lowers $f_{SD}$. As shown by Sort *et al*,[9] the vortex structure can be imprinted into the AF upon ZFC which then enhances and stabilizes the VS reversal. A lower $f_{SD}$ is then expected after ZFC relative to the FC procedure as we have observed. This relatively small change in $f_{SD}$ indicates that it is only those SD dots nearest the SD-VS phase boundary that now reverse as VS entities after ZFC.

In summary, we have investigated the interplay between geometric confinement and exchange bias in Fe/FeF$_2$ nanodots which contain a mixture of SD and VS reversal modes. The use of a FORC technique allows for the behavior of each reversal mode to be deconvoluted from the major loop. After FC, the SD dots show a simple loop shift whereas reversal of the VS dots is asymmetric, i.e. the nucleation fields exhibit a larger EB than the annihilation fields. These results agree with both Monte Carlo simulations and analytical calculations, and confirm the existence of LIDWs along the depth of the FM and tilted vortex cores. The macroscopic exchange bias extracted from conventional major loop measurements is found to be a weighted average of the local exchange bias experienced by the SD and VS nanodots. Moreover, the fraction of dots experiencing a given reversal mode depends on the cooling procedure, consistent with imprinting the vortex structure into the AF. These results unravel the crucial role played by local interfacial exchange on exchange bias and demonstrate a technique to deconvolute the behavior of different reversal modes across a macroscopic array of nanomagnets.

This work has been supported by the NSF (DMR-1008791, ECCS-0925626), the AFOSR (grant FA9550-10-1-0409) and the Texas A&M University-CONACyT collaborative research program.

**Figure Captions**

Fig. 1. Major hysteresis loops measured at 40 K, after separate field cooling (FC) and zero field cooling (ZFC) procedures, in comparison with that measured at room temperature (RT). Inset shows a scanning electron micrograph of 67nm unbiased Fe nanodots.

Fig. 2. Families of FORC's, whose starting points are represented by black dots, (a) at RT and (c) after FC to 40K. The corresponding FORC distributions are shown as contour plots in (b) and (d), respectively. The circled regions in (b) and (d) indicate the highly irreversible processes associated with the VS nucleation and annihilation while the dotted black box highlights the SD reversal mode. (e) Schematic hysteresis loop for unbiased vortices with nucleation field $H_N$ and annihilation field $H_A$ labeled. The inset shows an exchange biased loop. (f) Schematic FORC distribution indicating the relationship between the three primary FORC peaks in the ($H$, $H_R$) coordinate system and the nucleation/annihilation fields shown in (e).

Fig. 3. Projection of the FORC distribution along the $H_B$-axis, $dM/dH_B$, for the SD reversal mode at RT, after FC, and after ZFC. The shift of the peak along the $H_B$-axis indicates the exchange bias of the SD dots.

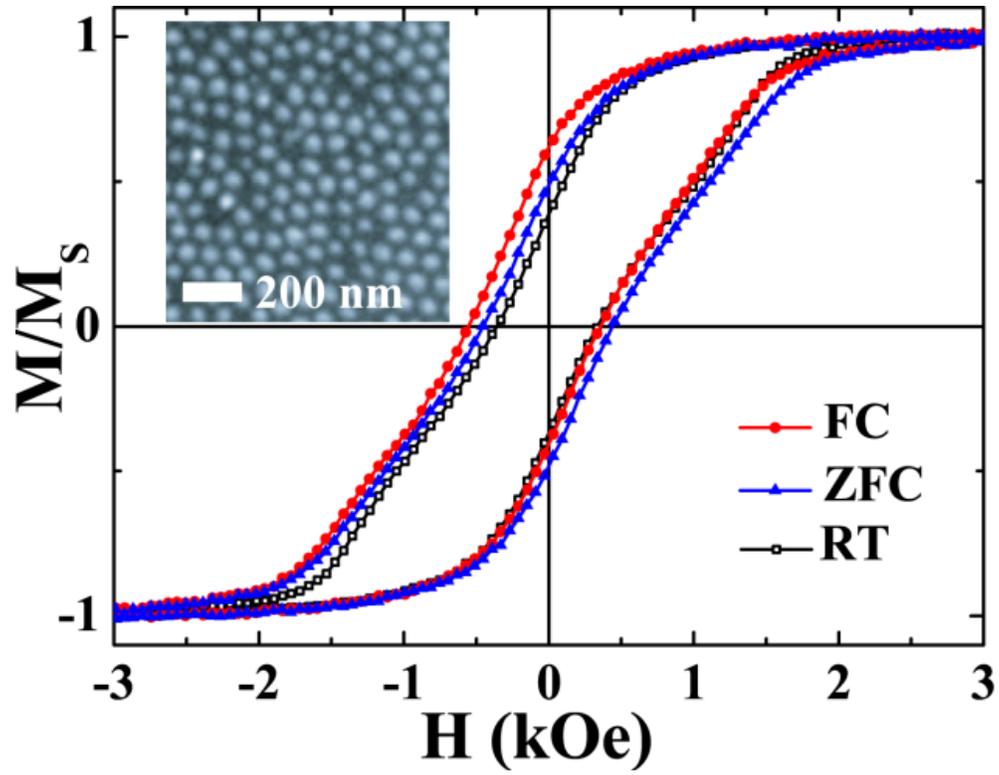

**Fig. 1. Dumas et al.**

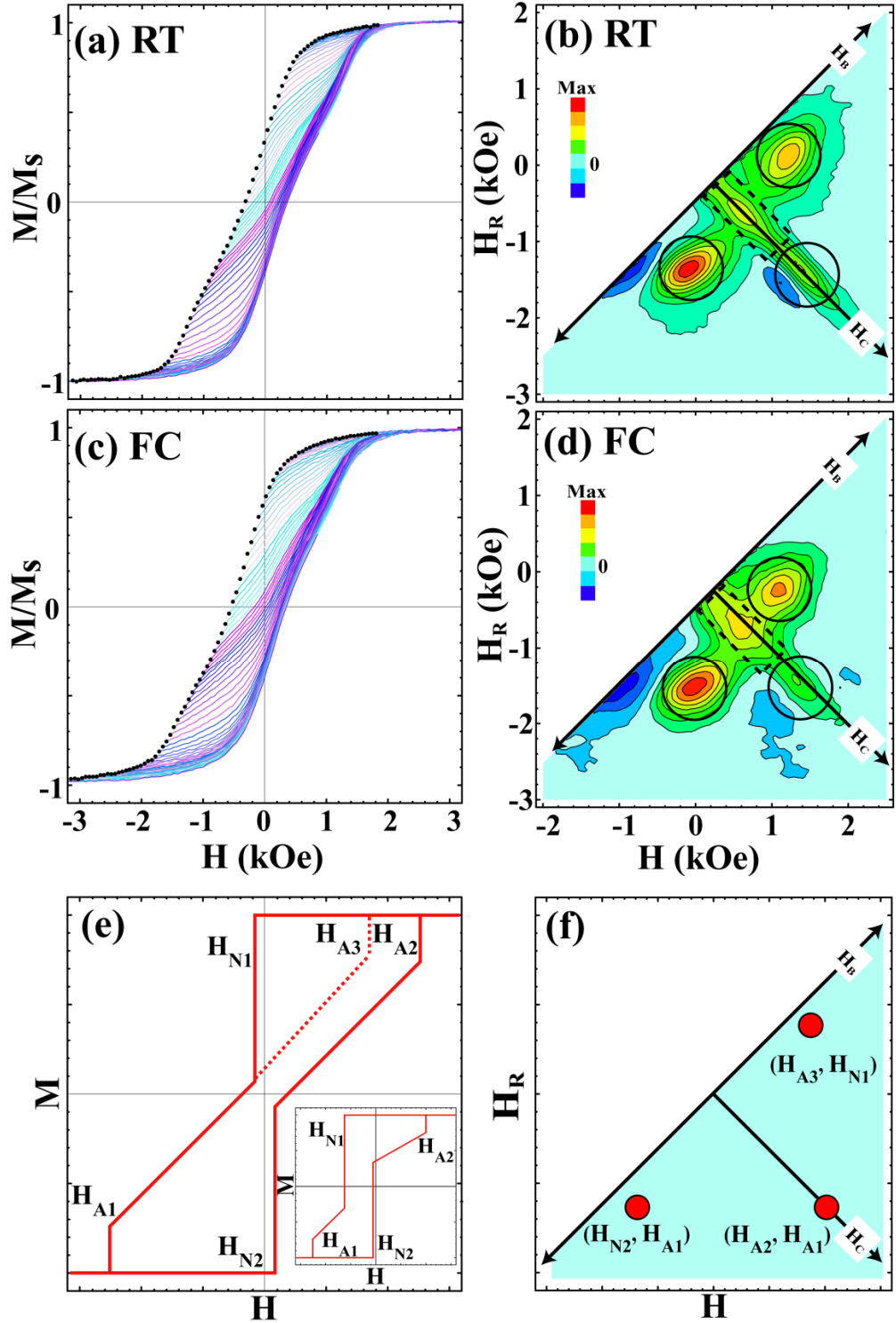

Fig. 2. Dumas et al.

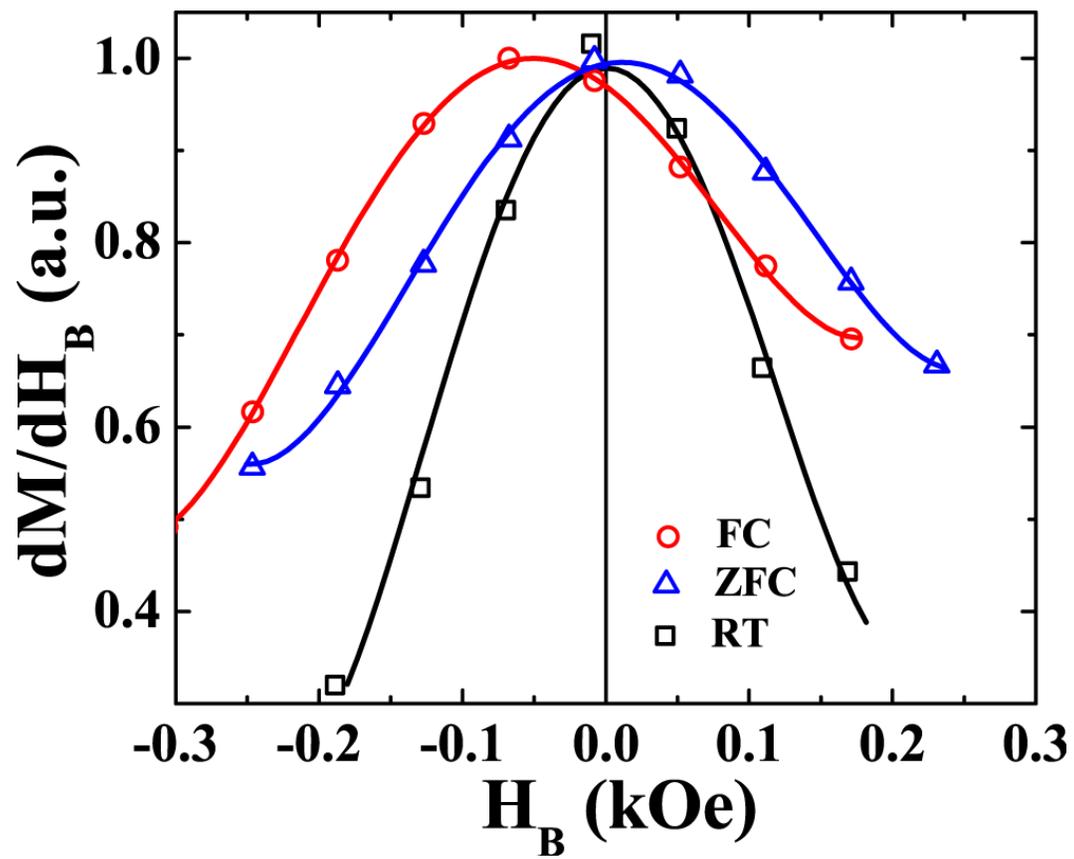

Fig. 3. Dumas et al.